\documentclass[aps,showpacs,twocolumn,floats]{revtex4}
\usepackage{graphicx}

\newcommand{\br}{{\bf r}}
\newcommand{\bk}{{\bf k}}

\newcommand{\bq}{{\bf q}}
\newcommand{\bp}{{\bf p}}

\begin{document}

\title{ Massless Dirac Fermions in  a Square Optical Lattice }
\author{Jing-Min Hou$^{1}$}
\email{jmhou@seu.edu.cn}
\author{Wen-Xing Yang$^{1,3}$}
\author{Xiong-Jun Liu$^2$}
\affiliation{$^1$Department of Physics, Southeast University,
Nanjing, 211189, China\\ $^2$ Department of Physics,
Texas A\&M University, College Station, Texas 77843-4242, USA\\
$^3$ Institute of Photonics Technologies, National Tsing-Hua
University, Hsinchu 300, Taiwan }
\date{January 5,  2009 }

\begin{abstract}
We propose a novel scheme to simulate and observe massless Dirac
fermions with cold atoms in a square optical lattice. A $U(1)$
adiabatic phase is created by two laser beams for the tunneling of
atoms between neighbor lattice sites. Properly adjusting the
tunneling phase, we find that the energy spectrum has conical points
in per Brillouin zone where band crossing occurs. Near  these
crossing points the quasiparticles and quasiholes can be considered
as massless Dirac fermions. Furthermore, the anisotropic effects of
massless Dirac fermions are obtained in the present square lattice
model. The Dirac fermions as well as the anisotropic behaviors
realizeded in our system can be experimentally detected with the
Bragg spectroscopy technique. \pacs{37.10.Jk, 03.75.Ss, 05.30.Fk}
\end{abstract}
\maketitle

\section{Introduction}

Realization of two-dimensional (2D) systems of massless Dirac
fermions is of great fundamental importance, in the light of many
exotic phenomena obtained in such systems, such as zero modes,
 fractional statistics, unconventional
Landau levels, parity anomaly, chirality, and anomalous quantum Hall
effects \cite{Semenoff,Jackiw,Haldane}. However, two-dimensional
massless Dirac field have not been observed untill the creation of
graphene, a monolayer of graphite \cite{Novoselov2,Novoselov3}.
Electrons in graphene, obeying a linear dispersion relation, behave
like massless Dirac fermions
\cite{Novoselov2,Novoselov3,Zhang,Li,Zheng,Gusynin,Hou,Jachiw2,Pachos}.

Besides graphene, physicists also make efforts to search for other
physical systems, e.g. patterned 2D electron gases \cite{Park} and
ultracold atoms in the honeycomb optical lattice
\cite{Zhu,Zhao,Shao,Wu} , to simulate massless Dirac fermions.
Realization of honeycomb optical lattice opens new possibility of
studying Dirac fermions in cold atoms which provide an extremely
clean environment and controllable fashion unique access to the
study of complex physics \cite{Jaksch,Greiner,Lewenstein}.
Nevertheless, all of the above systems require the hexagonal
symmetry. Then, it is very attractive to find a system without the
hexagonal symmetry to observe massless Dirac fermions.

Ultracold atom systems   provide an ideal platform to study many
interesting physics in condensed matters.   To investigate the
effects of gauge fields with ultracold atoms, several schemes have
been proposed to create an artificial Abelian gauge field
\cite{Jaksch2,Dum,Juzeliunas1,Juzeliunas2,Juzeliunas3,Gunter} or a
non-Abelian gauge field \cite{Osterloh,Ruseckas,Lu} for neutral
atoms with laser fields. Many effects have been studied for cold
atoms in an effective gauge field,   e.g., Stern-Gerlach effect for
chiral molecules\cite{Li2}, Double and negative
reflection\cite{Juzeliunas4}, Landau levels\cite{Jacob}, spin Hall
effect \cite{Liu,Zhu2}, induced spin-orbit coupling\cite{Liu2},
magnetic monopole\cite{Pietila}, spin field effect
transistors\cite{Vaishnav}. Furthermore, some groups have realized
the light-induced gauge fields in experiments\cite{Dutta, Lin}.

In this paper,  we propose a scheme to generate a staggered gauge
field with laser fields.  A 2D square lattice model under this
artificial gauge  field has a spectrum behaving like  massless Dirac
fermions. Furthermore, our lattice model does not have the hexagonal
symmetry. In our scheme, the energy bands of the system exhibit
degeneracy points where the conduction and valence bands intersect.
Near the these crossing points the dispersion relation is linearly
dependent on the momentum, say, is of the Dirac type. The present
scheme suggests a new direction to study Dirac fermions in the
optical lattice without the hexagonal symmetry.

\begin{figure}[ht]
\includegraphics[width=0.5\columnwidth]{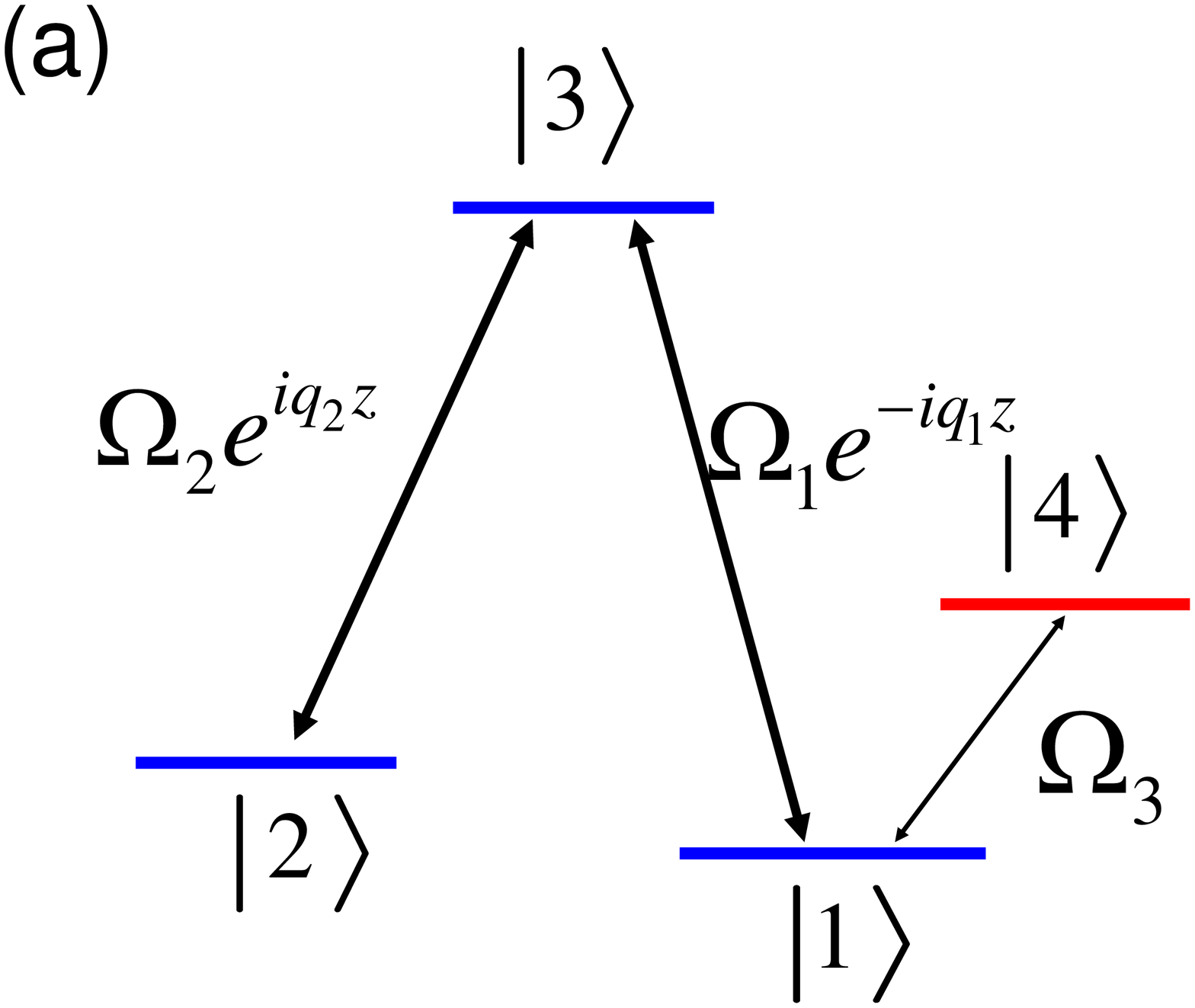}
\includegraphics[width=0.48\columnwidth]{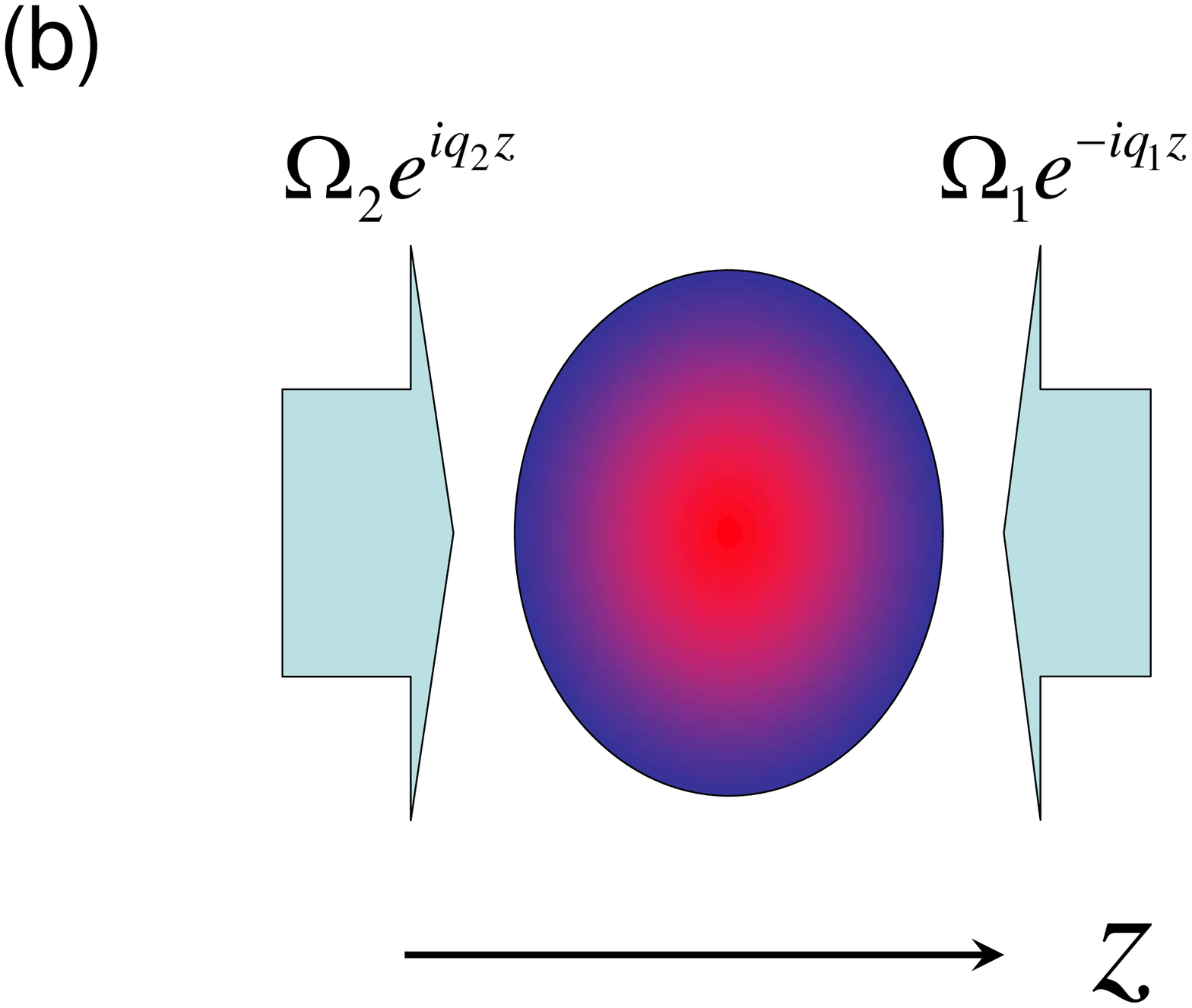}
\includegraphics[width=0.45\columnwidth]{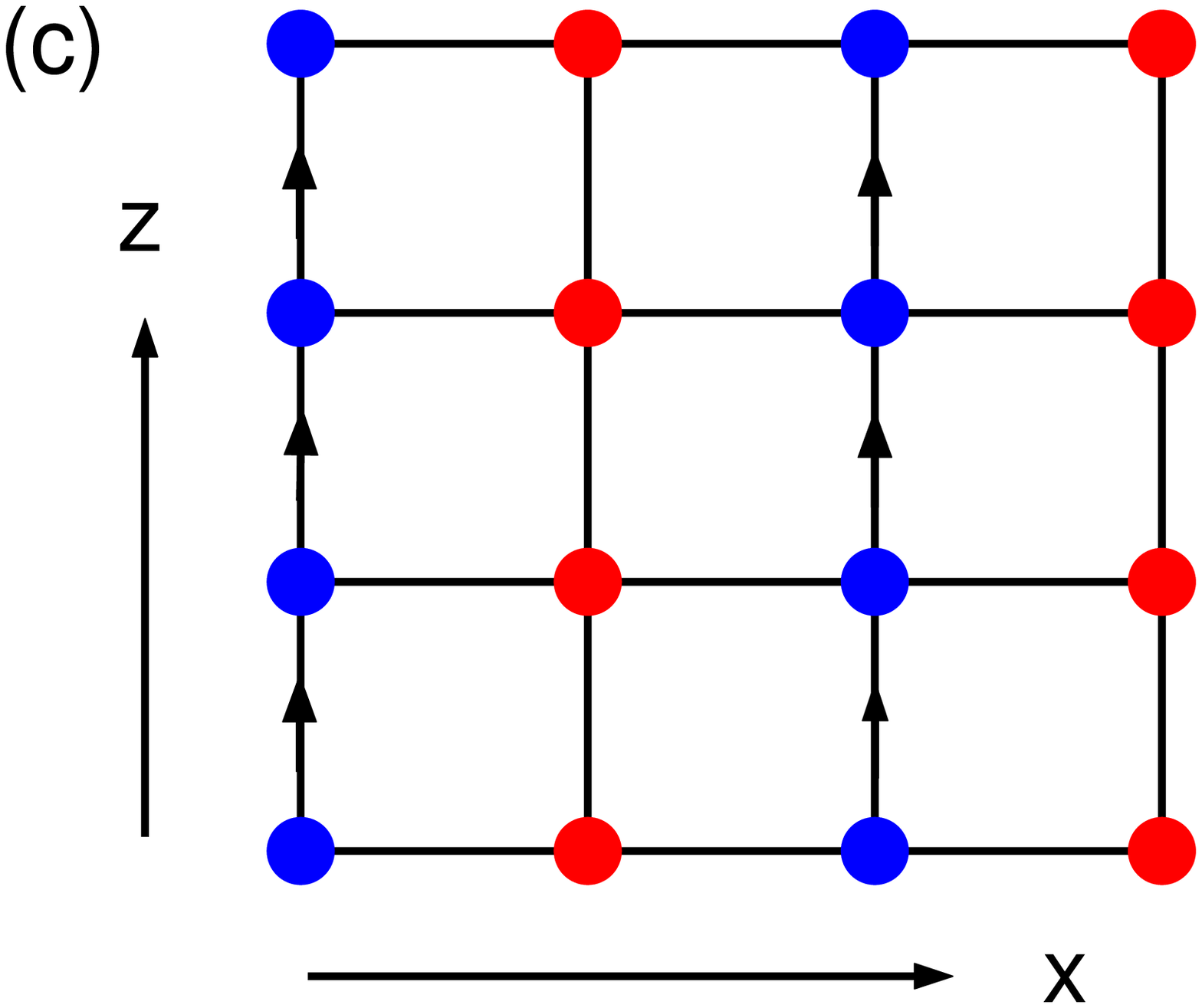}
\includegraphics[width=0.5\columnwidth]{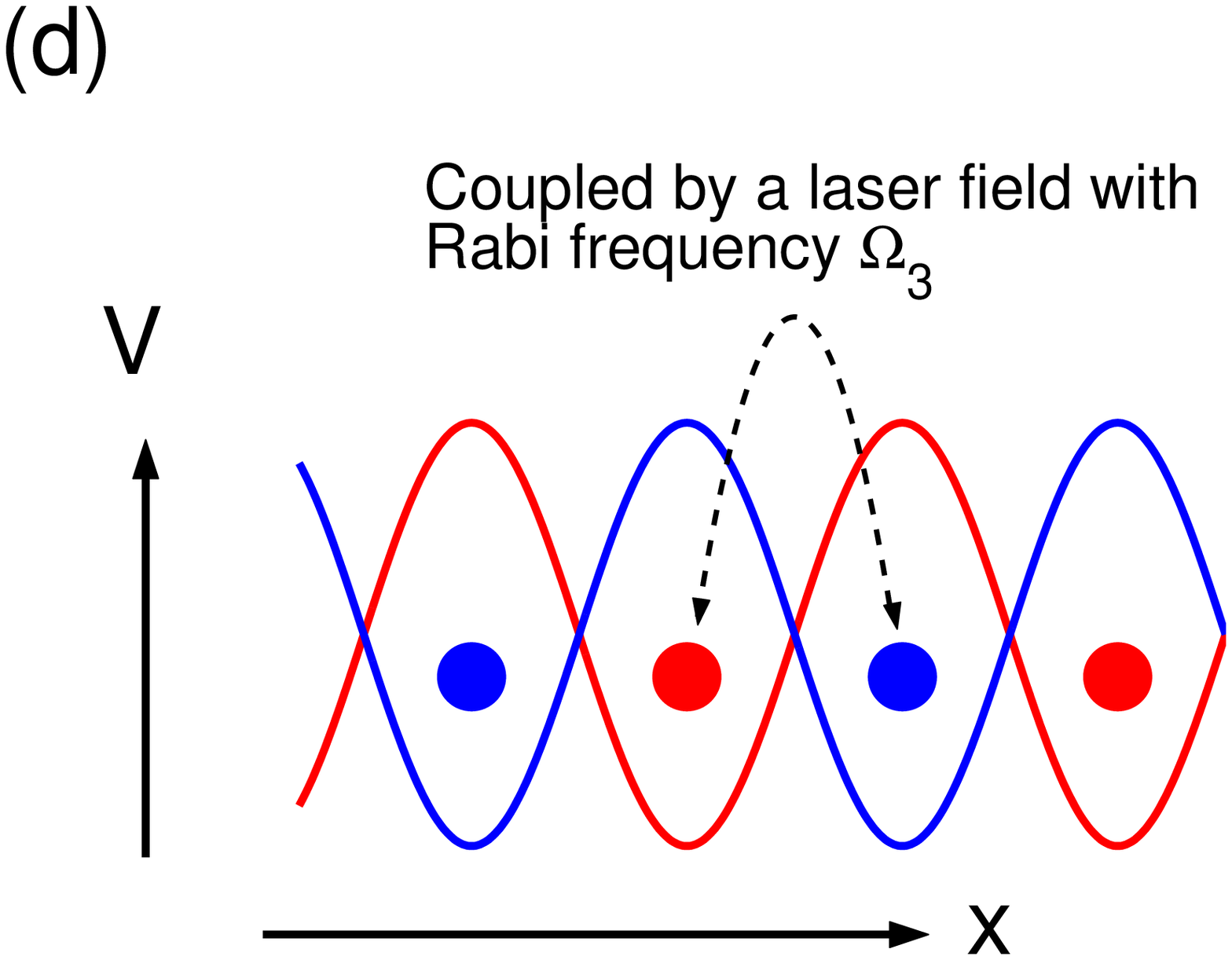}
\caption{(a) The atomic levels and the interactions between atoms
and laser fields. (b) Schematic representation of the experimental
setup with the two laser beams incident on the cloud of atoms. (c)
Schematic of the square optical lattice and the designed phase
factor (denoted by arrows). (d) The scheme of overlapping the two
state-selective optical lattices.}\label{lattice}
\end{figure}

\section{Model}

We consider a system of ultracold fermionic atoms with four levels
shown in FIG.\ref{lattice} (a). This atomic level configuration can
be experimentally realized with alkali atom $^6$Li \cite{Fuchs}. We
choose the atomic states $2S_{1/2}(F=1/2, m_F=1/2)$,
$2S_{1/2}(F=3/2, m_F=3/2)$, $2P_{1/2}(F=1/2, m_F=1/2)$ and
$2P_{1/2}(F=1/2, m_F=-1/2)$ as $|1\rangle, |2\rangle, |3\rangle$ and
$|4\rangle$, respectively.
 The cold atoms are trapped in two
state-selective optical potentials as shown in FIG.\ref{lattice} (c)
and (d). We assume that the states $|1\rangle$ and  $|2\rangle$ have
the same the state-selective optical potential, say sublattice $A$,
and $|4\rangle$  only perceives the other state-selective optical
potential, say sublattice $B$. Here, for convenience, we assume that
atoms in state $|3\rangle$  also perceive sublattice $A$. However,
this is unnecessary in our scheme, for the population of the quantum
state $|3\rangle$ is finally eliminated.  The two sublattices have
the lattice spacings $2l_x$ and $l_z$ in the $x$ and $z$ directions,
respectively. The two sublattices make up a 2D rectangular lattice
with the lattice spacings $l_x$ and $l_z$, especially a 2D square
lattice for $l_x=l_z$, when overlapping together as shown in
FIG.\ref{lattice} (c) and (d). Without loss of generality, we
suppose that atoms with internal states $|1\rangle $ and $|2\rangle$
are trapped in odd columns and ones with internal states $|4\rangle$
in even columns in the whole overlapped lattice.
  For convenience, we
assume that the 2D square lattice considered here is in the $x-z$
plane
 as shown in FIG.\ref{lattice} (c). Two additional laser beams
along the $y$ direction are added. When the potential barrier of the
optical lattice along the $y$ direction is high enough, the
tunneling along this direction between different planes is
suppressed seriously, then every layer is an independent 2D lattice
in $x-z$ plane.

Using $\{|1\rangle, |2\rangle,|3\rangle,|4\rangle\}$ as the basis,
the Hamiltonian of free ultracold fermions in the optical lattice
can be written in the second quantized form as follows,
\begin{eqnarray}
\hat H_{0}&=& \int d^2r
\hat\Psi^\dag\left(-\frac{\hbar^2}{2m}\nabla^2+V(\br)\right)\hat\Psi,
\label{H0}
\end{eqnarray}
where $\hat\Psi^\dag=(\hat\Psi_1^\dag, \hat\Psi_2^\dag,
\hat\Psi_3^\dag, \hat\Psi_4^\dag)$ and $\hat\Psi=(\hat\Psi_1,
\hat\Psi_2, \hat\Psi_3, \hat\Psi_4)^T$ ($T$ denotes the matrix
transposition)  with $\hat\Psi_{i}(\br)$ and
$\hat\Psi_{i}^\dag(\br)$ being field operators corresponding to
annihilating  and creating an atom with the internal quantum state
$|i\rangle \ (i=1,2,3,4)$ at coordinate position $\br$ respectively.
Here, $V({\bf r})$ is the trap potential matrix as
\begin{eqnarray}
V({\bf r})=\left(\matrix{V_A({\bf r})&0&0&0\cr 0&V_A({\bf r})&0&0\cr
0&0&V_A({\bf r})&0\cr 0&0&0&V_B({\bf r})}\right),
\end{eqnarray}
where $V_X(\br)(X=A,B) $ are the two state-selective periodic
potentials. The ground state $|1\rangle$ is coupled to the excited
state $|3\rangle$ via a  laser field with the corresponding Rabi
frequency $\Omega_1 e^{-iq_1z}$ and the state $|2\rangle$ is coupled
to the excited state $|3\rangle$ via a laser field with the
corresponding Rabi frequency $\Omega_2 e^{iq_2z}$ as shown in
FIG.\ref{lattice} (a) and (b).
 The corresponding light-atom interaction
Hamiltonian is,
\begin{eqnarray}
\hat H_1&=&\int d^2r \hat\Psi^\dag M\hat\Psi,
 \label{IH}
\end{eqnarray}
with
\begin{eqnarray}
M=\hbar\left(\matrix{0&0&\Omega_1e^{iq_1z}&0\cr
0&0&\Omega_2e^{-iq_2z}&0\cr
\Omega_1e^{-iq_1z}&\Omega_2e^{iq_2z}&0&0\cr 0&0&0&0}\right),
\end{eqnarray}
 where $\Omega_j (j=1,2)$ are the Rabi frequencies.
 Additionally, the quantum state $|1\rangle$ is coupled to the
 quantum
state $|4\rangle$ via a laser field propagating in the $y$ direction
with Rabi frequency $\Omega_3e^{iq_3y}$. Because $e^{iq_3y}$ is a
constant in the $x-z$ plane, we can omit this phase factor by
supposing the two-dimensional lattice on the $y=0$ plane. The
corresponding interaction Hamiltonian is
\begin{eqnarray}
 \hat H_2=\int
d^2r
 \hat\Psi^\dag N\hat\Psi,
 \end{eqnarray}
 with
\begin{eqnarray}
N=\hbar\left(\matrix{0&0&0&\Omega_3\cr 0&0&0&0\cr 0&0&0&0\cr\Omega_3
&0&0&0}\right).
\end{eqnarray}
  The total Hamiltonian can
be written as $\hat H=\hat H_{0}+\hat H_1+\hat H_2$.

 The Hamiltonian (\ref{IH}) can be diagonalized by the matrix,
\begin{eqnarray}
U=\left(\matrix{\cos\theta& -\sin\theta e^{iqz}&0&0
\cr\frac{\sqrt{2}}{2}\sin\theta
e^{-iqz}&\frac{\sqrt{2}}{2}\cos\theta
&-\frac{\sqrt{2}}{2}e^{-iq_2z}&0\cr \frac{\sqrt{2}}{2}\sin\theta
e^{-iqz}&\frac{\sqrt{2}}{2}\cos\theta&\frac{\sqrt{2}}{2}e^{-iq_2z}&0\cr
0&0&0&1}\right),
\end{eqnarray}
where $q=q_1+q_2$ and $\tan\theta=|\Omega_1|/|\Omega_2|$.
Correspondingly, we obtain the dressed states as
\begin{eqnarray}
|\chi_1\rangle&=&\cos\theta| 1\rangle -\sin\theta e^{iqz}| 2\rangle,\\
 |\chi_2\rangle&=&  \frac{\sqrt{2}}{2}\sin\theta
e^{-iqz}| 1\rangle+\frac{\sqrt{2}}{2}\cos\theta| 2\rangle-\frac{\sqrt{2}}{2} e^{-iq_2z}| 3\rangle,\\
  |\chi_3\rangle&=&\frac{\sqrt{2}}{2}\sin\theta
e^{-iqz}| 1\rangle+\frac{\sqrt{2}}{2}\cos\theta|
2\rangle+\frac{\sqrt{2}}{2} e^{-iq_2z}| 3\rangle,\\
|\chi_4\rangle&=&|4\rangle,
\end{eqnarray}
with the energy eigenvalues  $ E_i=(0,-\hbar\Omega,\hbar\Omega,0)$
with $\Omega=\sqrt{|\Omega_1|^2+|\Omega_2|^2}$. Here, the state
$|\chi_1\rangle$ is a so-called dark state, which does not contain
the component of the excited atomic state $|3\rangle$, and
$|\chi_2\rangle, |\chi_3\rangle$ are   bright states. In the dressed
state basis $\{ |\chi_1\rangle, |\chi_2\rangle, |\chi_3\rangle,
|\chi_4\rangle\}$, the  vector field operator can be written as $
\hat\Phi=({\hat\Phi_{1}, \hat\Phi_{2}}, \hat\Phi_{3},
\hat\Phi_{4})^T=U({\hat\Psi_{1}, \hat\Psi_{2}},\hat\Psi_{3},
\hat\Psi_{4})^T$, where $\hat\Phi_{j} (j=1,2,3,4)$ represent
destructing an atom in the dressed state $|\chi_j\rangle
(j=1,2,3,4)$. Thus, the Hamiltonian can be rewritten as
\begin{eqnarray}
\hat{H}=\int d^2r
\hat\Phi^\dag\left[\frac{1}{2m}(-i\hbar\nabla-\tilde{\bf
A})^2+\tilde{V}(\br)+\tilde{N}\right]\hat\Phi,
\end{eqnarray}
where $\tilde{\bf A}=i\hbar U\nabla U^\dag$, $\tilde{V}({\bf
r})=UV({\bf r})U^\dag +UMU^\dag+\frac{\hbar^2}{2m}[(U\nabla
U^\dag)^2+\nabla U\cdot\nabla U^\dag]$ and $\tilde{N}=UNU^\dag$. We
straightforwardly calculate these matrices and obtain,
\begin{widetext}
\begin{eqnarray}
\tilde{\bf A} &=&-\hbar{\bf
e}_z\left(\matrix{-q\sin^2\theta&\frac{\sqrt{2}}{2}q\sin\theta\cos\theta
e^{iqz}&\frac{\sqrt{2}}{2}q\sin\theta\cos\theta e^{iqz}&0\cr
\frac{\sqrt{2}}{2}q\sin\theta\cos\theta
e^{-iqz}&\frac{1}{2}q\sin^2\theta+\frac{1}{2}q_2&\frac{1}{2}q\sin^2\theta-\frac{1}{2}q_2&0\cr
\frac{\sqrt{2}}{2}q\sin\theta\cos\theta
e^{-iqz}&\frac{1}{2}q\sin^2\theta-\frac{1}{2}q_2&\frac{1}{2}q\sin^2\theta+\frac{1}{2}q_2&0\cr
0&0&0&0}\right),
\end{eqnarray}
and
\begin{eqnarray} \tilde{V}({\bf r})
 &=&\left(\matrix{V_A({\bf r})&0&0&0\cr
0&V_A({\bf r})-\hbar\Omega&0&0\cr 0&0&V_A({\bf r})+\hbar\Omega&0\cr
0&0&0&V_B({\bf r})}\right),
\end{eqnarray}
and
\begin{eqnarray}
\tilde{N} &=&\hbar\left(\matrix{0&0&0&\Omega_3\cos\theta\cr
0&0&0&\frac{\sqrt{2}}{2}\Omega_3\sin\theta e^{-iqz}\cr
0&0&0&\frac{\sqrt{2}}{2}\Omega_3\sin\theta e^{-iqz}\cr
\Omega_3\cos\theta&\frac{\sqrt{2}}{2}\Omega_3\sin\theta
e^{iqz}&\frac{\sqrt{2}}{2}\Omega_3\sin\theta e^{iqz}&0}\right).
\end{eqnarray}
\end{widetext}

In our scheme, we only consider the atoms  in the dressed states
$|\chi_1\rangle$ and $|\chi_4\rangle$. Thus, we have to
adiabatically eliminate the populations of the dressed states
$|\chi_2\rangle$ and $|\chi_3\rangle$ and to avoid the atoms
decaying  into these two dressed states. This can be realized in the
steps. First, we start with the atoms in  the atomic state
$|1\rangle$ and $\Omega_1=0$, $\Omega_3=0$ with $\Omega_2$ finite,
then slowly turn $\Omega_1$, we will end up with the atoms in the
dressed state $|\chi_1\rangle$ \cite{Scully}. During this process,
the variation of $\Omega_1$ is slow enough to satisfy the adiabatic
condition $|\langle \chi_j|\partial/\partial t|\chi_1\rangle|\ll
|E_j-E_1|/\hbar=\Omega$ with $j=2,3$ \cite{Messiah,jmhou}. In the
second step, we adiabatically turn the Rabi frequency $\Omega_3$ on,
we will end up with atoms in the dressed states $|\chi_1\rangle$ and
$|\chi_4\rangle$. To avoid the atoms decaying into the dressed
states $|\chi_2\rangle$ and $|\chi_3\rangle$, the adiabatic
conditions  $\frac{1}{2m}|\tilde{\bf
A}_{j1}|^2=\frac{\hbar^2}{2m}q^2\sin^2\theta\cos^2\theta\ll
|E_{j}-E_{1}|=\hbar\Omega$ and
$|\tilde{N}_{j4}|=\frac{\sqrt{2}}{2}\hbar\Omega_3\sin\theta\ll
|E_{j}-E_4|=\hbar\Omega$ for $j=2,3$ are satisfied. This is to say,
the off-diagonal elements of the Hamiltonian are small enough to
avoid the atoms decaying into the dressed states $|\chi_2\rangle$
and $|\chi_3\rangle$.

 Since the atoms are only in the dressed states  $|\chi_{1}\rangle$  and $|\chi_{4}\rangle$, we  consider the reduced  space with
 the dressed state basis $\{|\chi_1\rangle, |\chi_4\rangle\}$.
  Therefore, the total Hamiltonian   can be reduced to
\begin{eqnarray}
\hat H&=&\int d^2r
\hat\Phi_{1}^\dag\left[\frac{1}{2m}(-i\hbar\nabla-{\bf
{A}})^2+{V}_{A}(\br)\right]\hat\Phi_{1} \nonumber\\
&+&\int d^2r \hat\Phi_{4}^\dag\left[-\frac{\hbar^2}{2m}\nabla^2+{V}_{B}(\br)\right]\hat\Phi_{4} \nonumber\\
 &+&\hbar\Omega_e\int d^2r
\left(\hat\Phi_{4}^\dag\hat\Phi_{1}+\hat\Phi_{1}^\dag\hat\Phi_{4}\right),\label{eh}
\end{eqnarray}
where $\Omega_e=\tilde{N}_{14}/\hbar=\Omega_3\cos\theta$ and the
$U(1)$ adiabatic gauge potential ${\bf A}=\tilde{\bf A}_{11}=\hbar
q\sin^2\theta{\bf e}_z$.

\section{Massless Dirac fermions}

Taking the tight-binding limit, we can superpose the Bloch states to
get Wannier functions $w_a(\br-\br_i)$ and $w_b(\br-\br_j)$ for
sublattice $A$ and $B$, respectively. In the present case, we can
expand the field operator in the lowest band Wannier functions as, $
\hat\Phi_{1}(\br)=\sum_{m(odd),n}\hat{a}_{m,n}e^{\frac{i}{\hbar}\int_0^{{\bf
r}_{mn}}{\bf A}\cdot d{\bf r}}w_a(\br-\br_{mn})$ and $
\hat\Phi_{4}(\br)=\sum_{m(even),n}\hat{b}_{m,n}w_b(\br-\br_{mn})$.
Substituting the above expression into Eq.(\ref{eh}), we can rewrite
the Hamiltonian as follows,
\begin{eqnarray}
\hat H&=&-\sum_{(m(odd),n)}
[t_b\hat b^\dag_{m+1,n+1}\hat b_{m+1}+t_ae^{i\gamma}\hat a^\dag_{m,n+1}\hat a_{m,n}\nonumber\\
&&+2t_1\hat a^\dag_{m,n}\hat b_{m+1,n}+{\rm H.c.}]+\hat H_K,
\label{ehr}
\end{eqnarray}
with
$H_K=\epsilon_a\sum_{(m(odd),n)}\hat{a}_{m,n}^\dag\hat{a}_{m,n}+\epsilon_b\sum_{(m(even),n)}\hat{b}_{m,n}^\dag\hat{b}_{m,n}$.
Here, the parameters have the following forms: $t_a=\int
d^2rw^*_a(\br-\br_{m,n+1})(-\hbar^2\nabla^2/2m+V_A)w_a(\br-\br_{mn})$,
$t_b=\int
d^2rw^*_b(\br-\br_{m,n+1})(-\hbar^2\nabla^2/2m+V_B)w_b(\br-\br_{mn})$,
$t_1=\Omega_e\int d^2rw^*_b(\br-\br_{m+1,n})w_a(\br-\br_{mn})$,
$\epsilon_a=\int
d^2rw^*_a(\br-\br_{m,n})(-\hbar^2\nabla^2/2m+V_A)w_a(\br-\br_{mn})$,
 $\epsilon_b=\int
d^2rw^*_b(\br-\br_{m,n})(-\hbar^2\nabla^2/2m+V_B)w_b(\br-\br_{mn})$
and  $\gamma=2\pi\sin^2\theta \hbar q l_z$ is the phase resulted
from the adiabatic gauge potential.  In our scheme, we consider
$\epsilon_a=\epsilon_b$, so $\hat H_K$ in Eq. (\ref{ehr}) can be
dropped out as a constant term, which does not affect the physics
considered here.

\begin{figure}[ht]
\includegraphics[width=0.45\columnwidth]{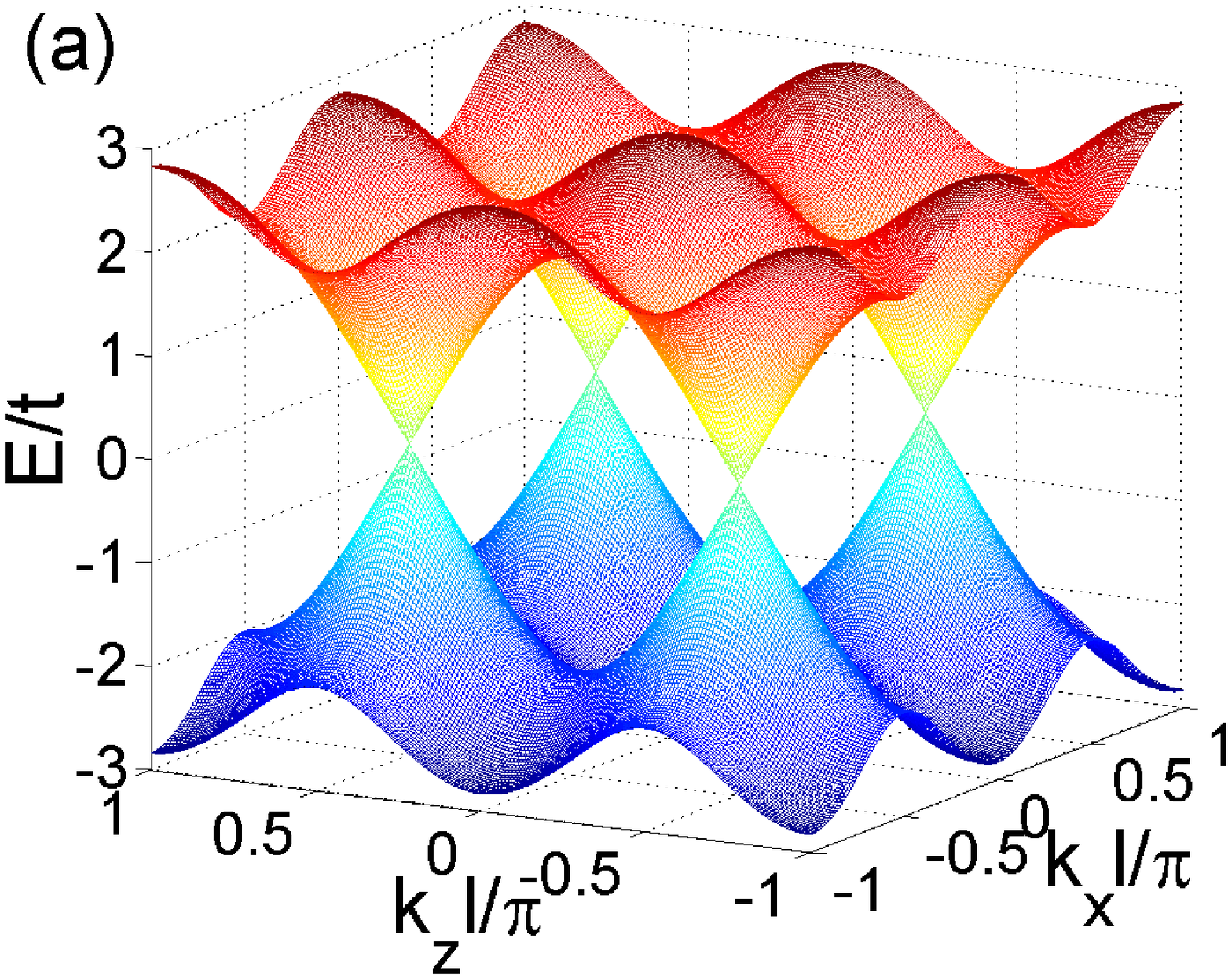}
\hspace{0.4cm}
\includegraphics[width=0.28\columnwidth]{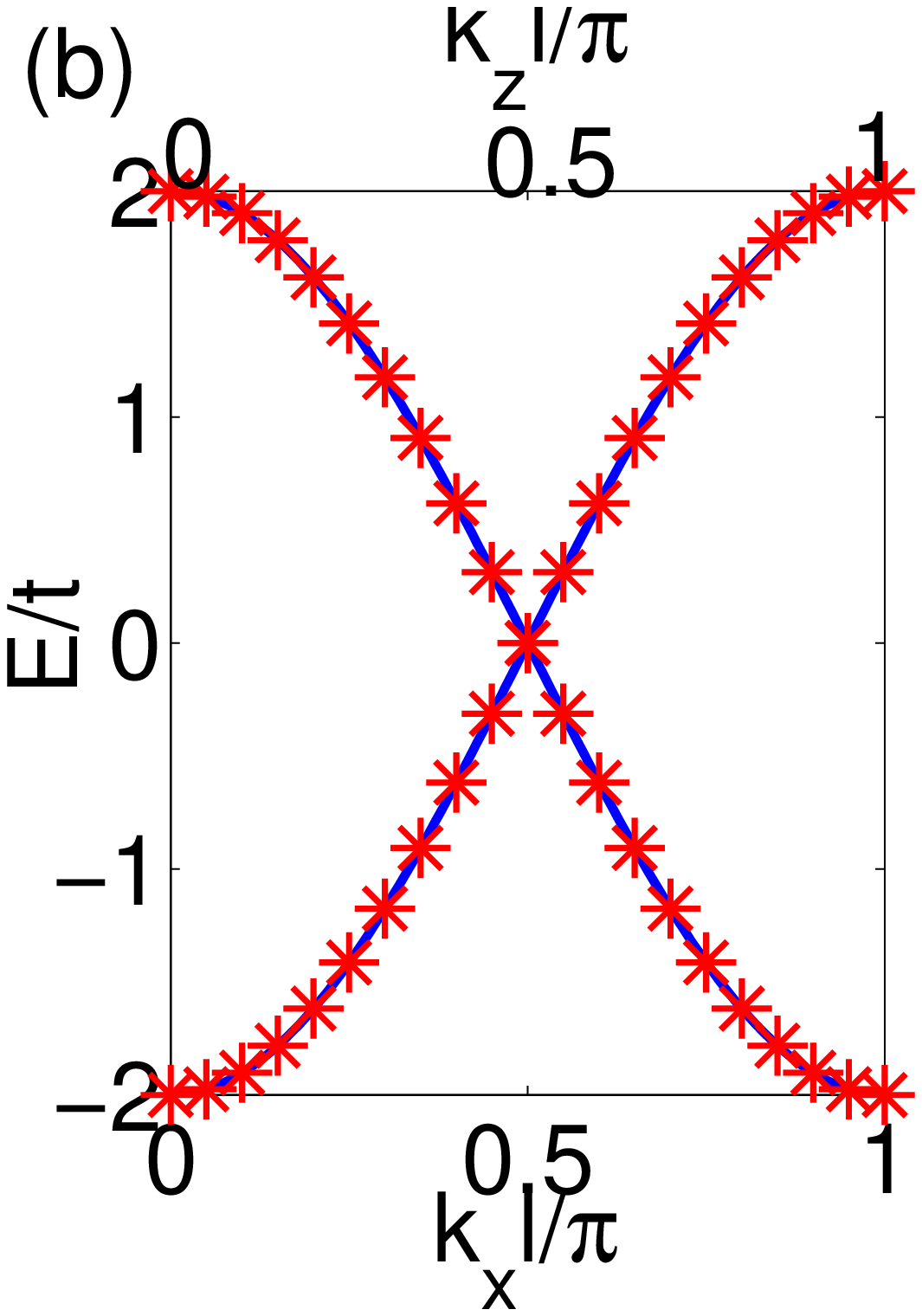}
\includegraphics[width=0.45\columnwidth]{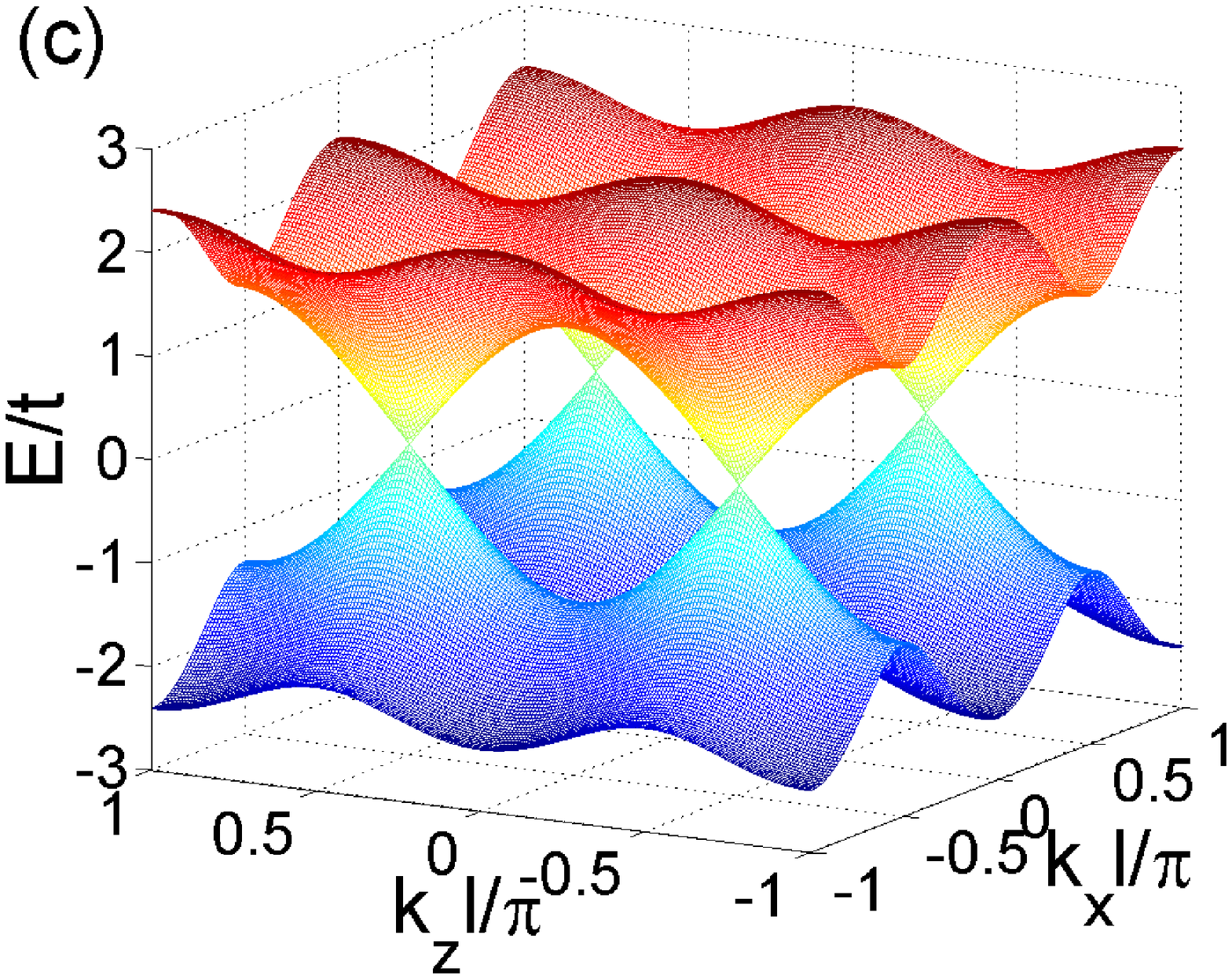}
\hspace{0.4cm}
\includegraphics[width=0.28\columnwidth]{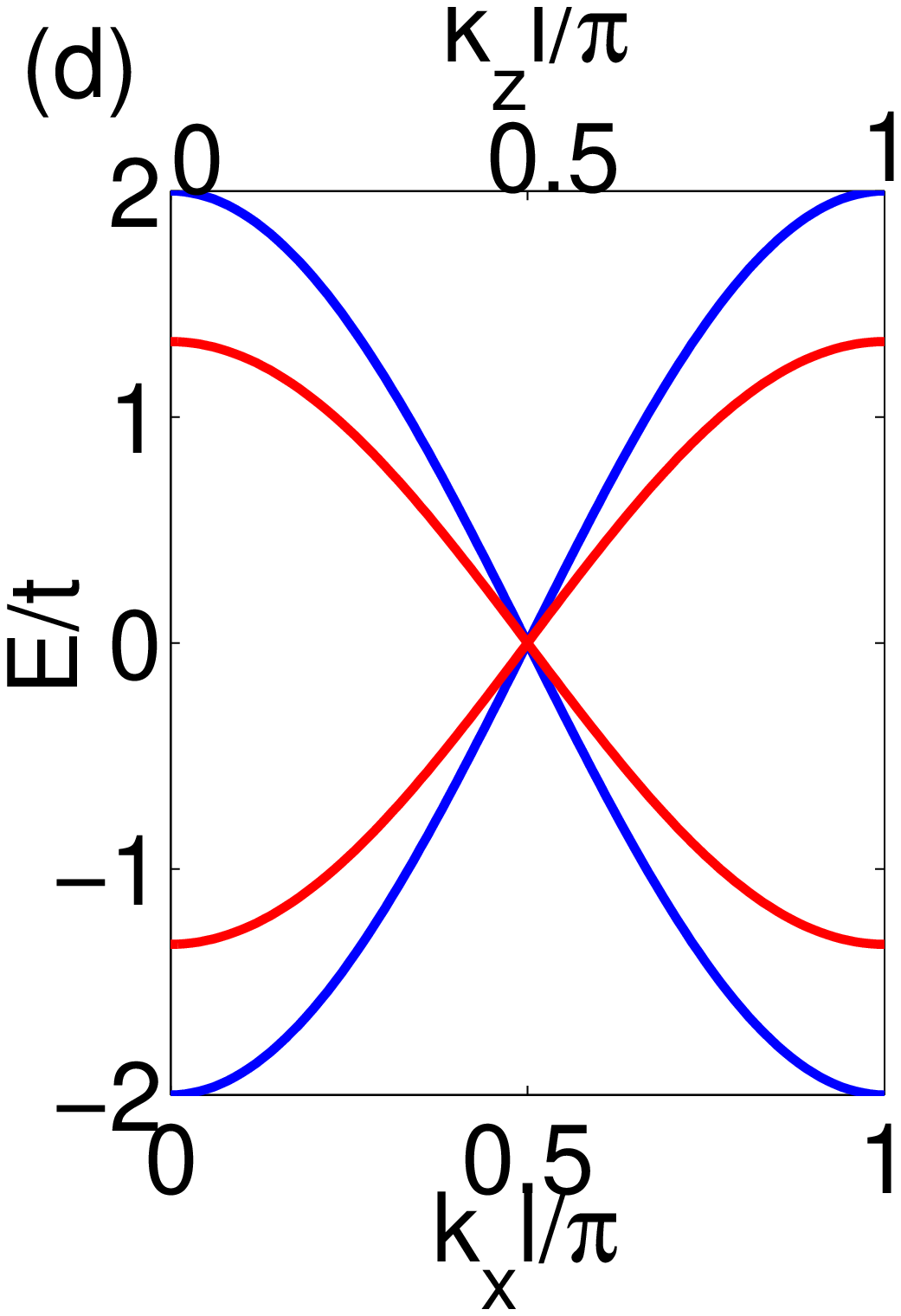}
\includegraphics[width=0.45\columnwidth]{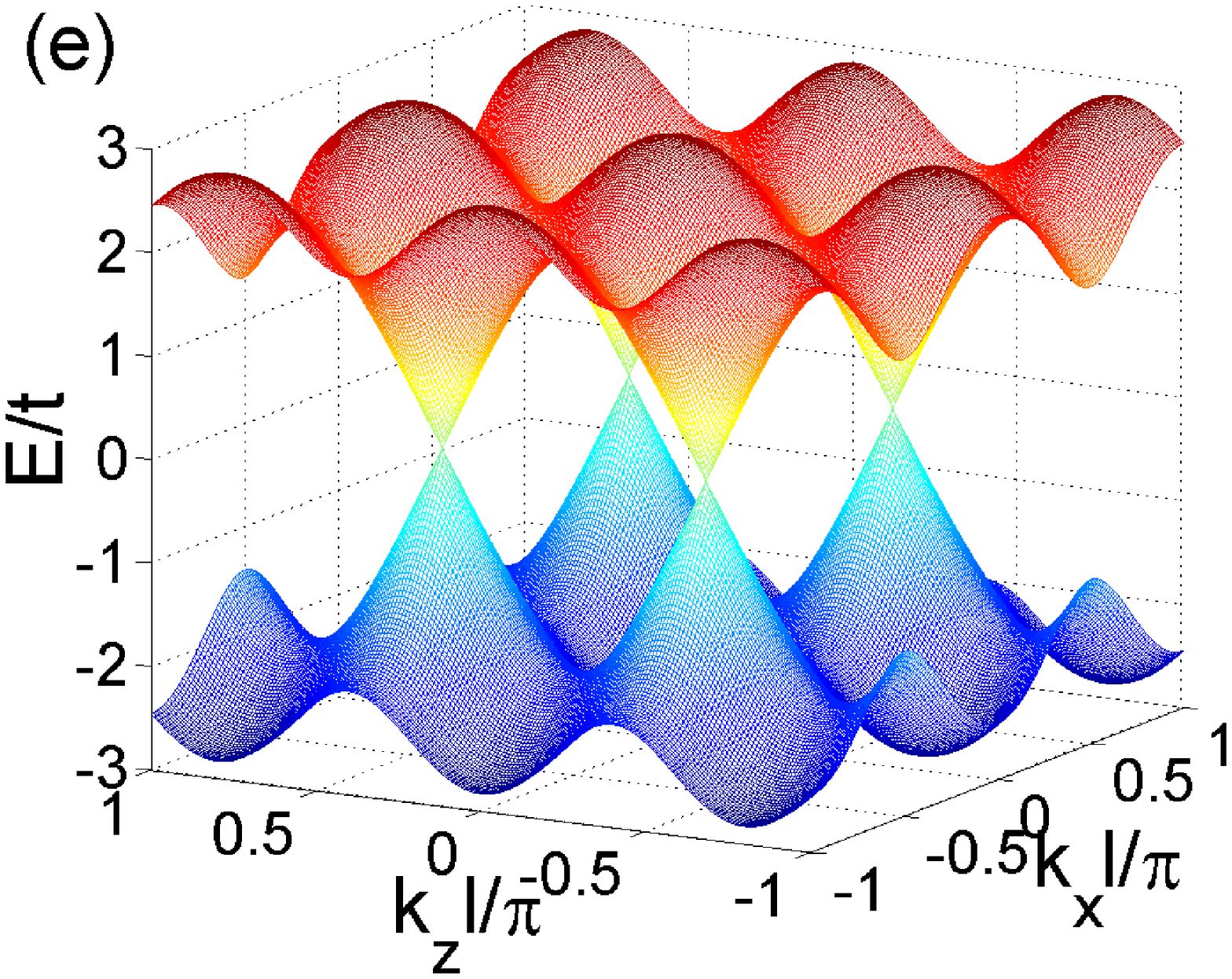}
\hspace{0.4cm}
\includegraphics[width=0.28\columnwidth]{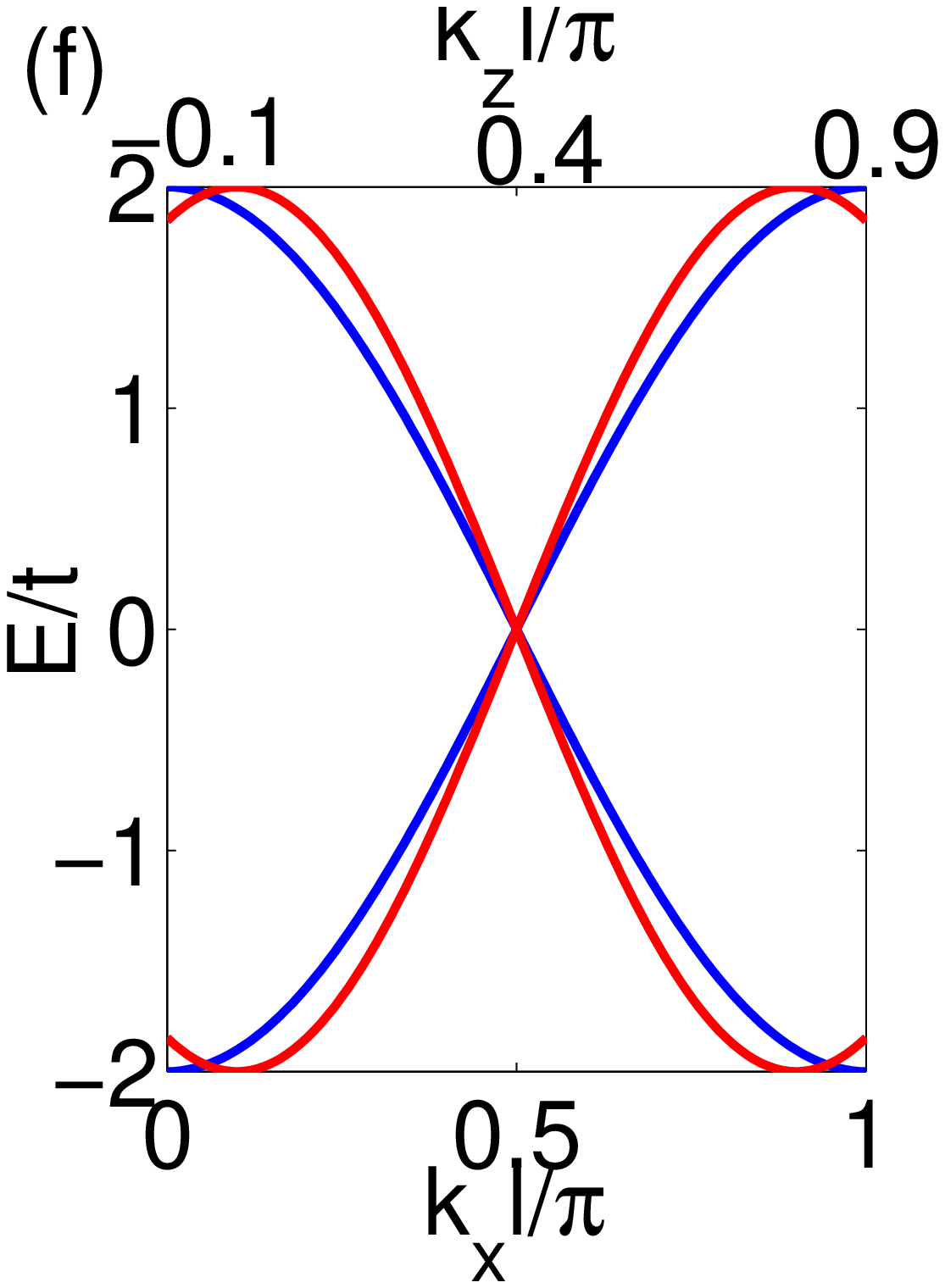}
\includegraphics[width=0.45\columnwidth]{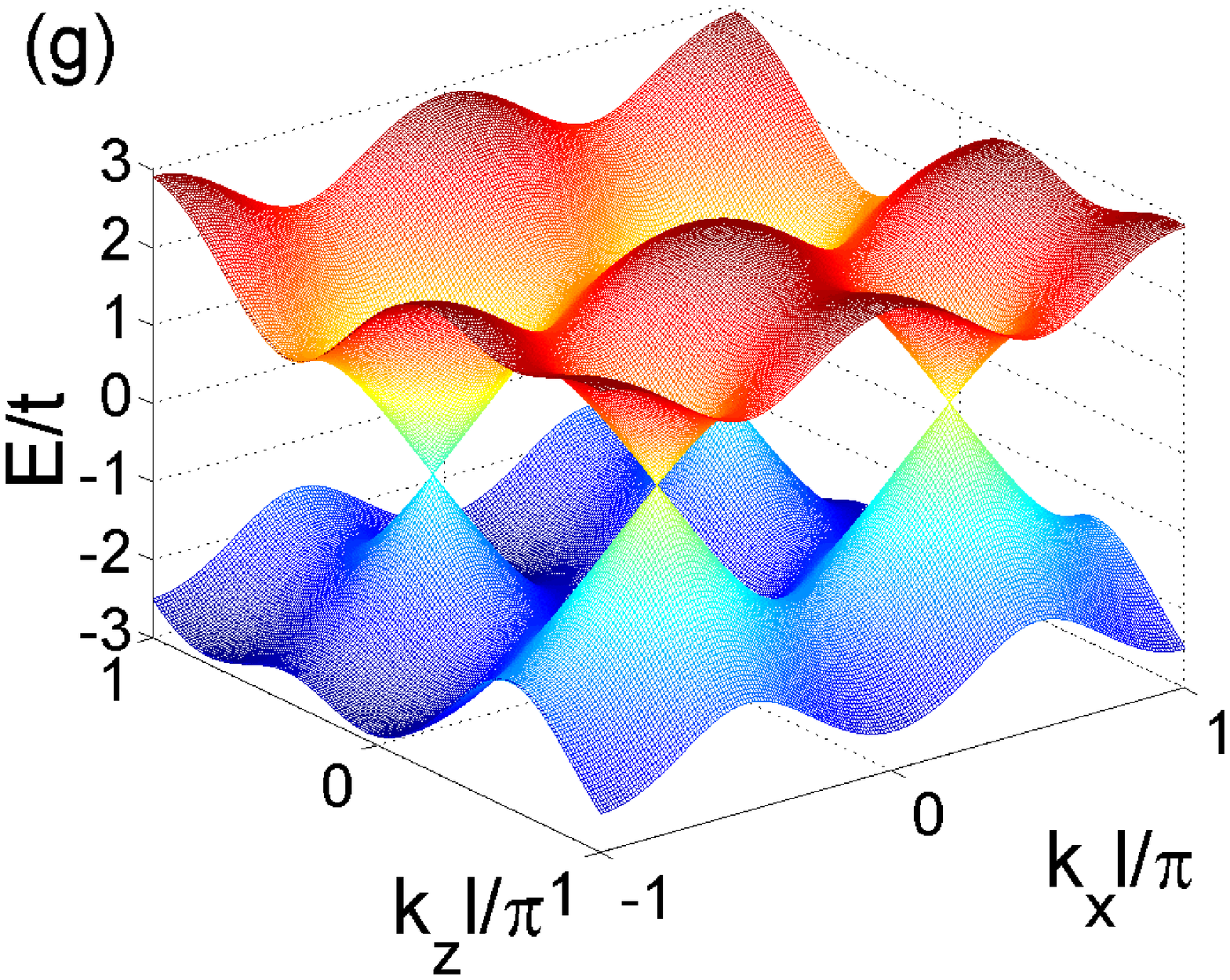}
\hspace{0.4cm}
\includegraphics[width=0.28\columnwidth]{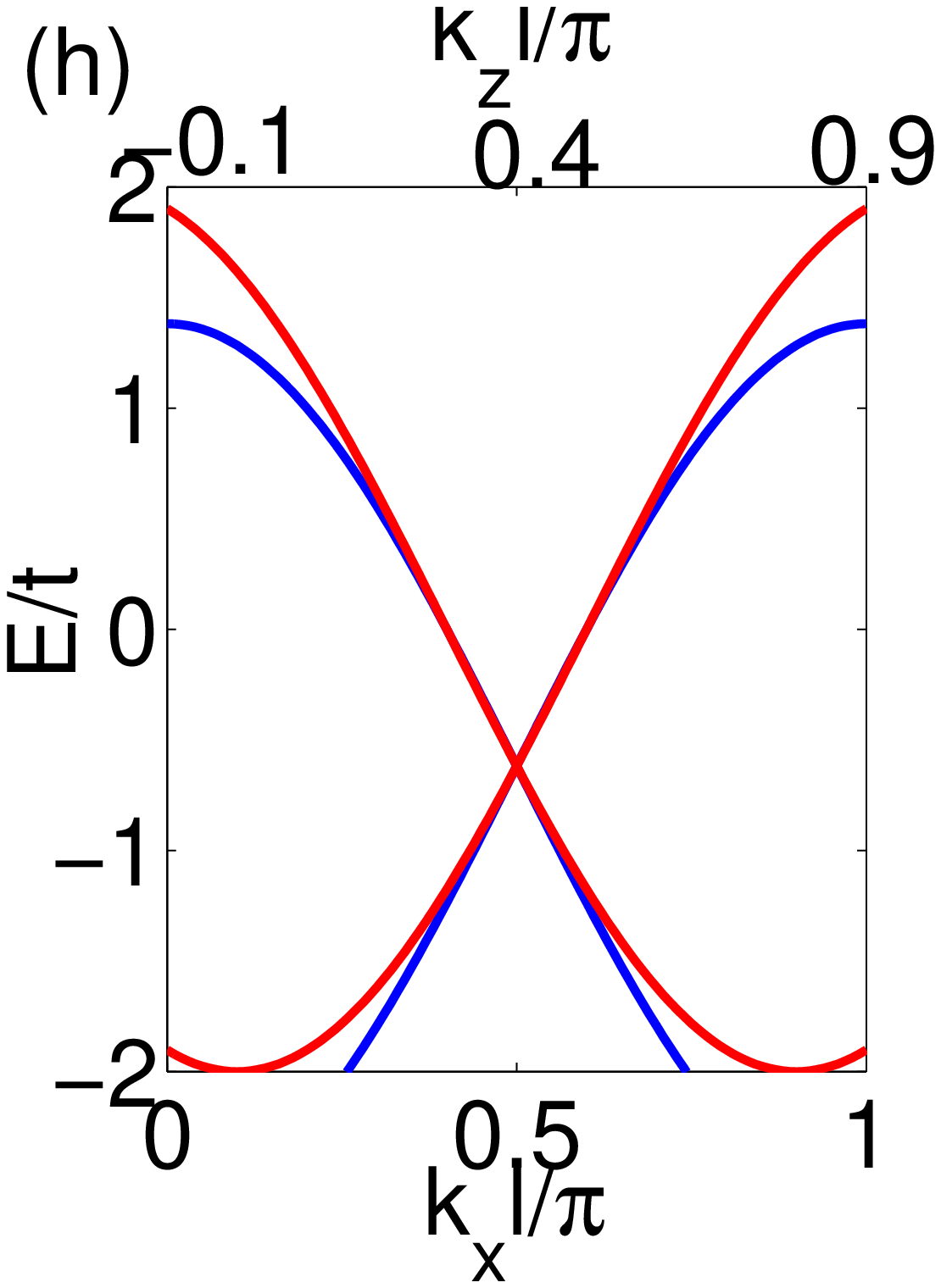}
 \caption{Energy dispersion for cold fermionic atoms in
a square optical lattice. (a)   shows the energy dispersion  and (b)
represents the profiles of the energy dispersion with $k_z=\pi/2l$
(blue line) and $k_x=\pi/2l$(red star), for the ideal case
$t_a=t_b=t_1=t$, $l_x=l_z=l$, $\gamma=\pi$. (c)  shows the energy
dispersion  and (d) represents the profiles of the energy dispersion
with $k_z=\pi/2l$ (blue line) and $k_x=\pi/2l$ (red line), for the
anisotropic case $t_a=t_b=2t_1/3=2t/3$,  $l_x=l_z=l$, $\gamma=\pi$.
(e) shows the energy dispersion  and (f) represents the profiles of
the energy dispersion with $k_z=2\pi/5l$ (blue line) and
$k_x=\pi/2l$(red line), for the anisotropic case $t_a=t_b=t_1=t$,
$l_z=5l_x/4=5l/4$, $\gamma=\pi$. (g) shows the energy dispersion and
(h) represents the profiles of the energy dispersion with
$k_z=2\pi/5l$
 (blue line) and $k_x=\pi/2l$ (red line), for the anisotropic case
with $t_a=t_b=t_1=t$, $l_x=l_z=l$, $\gamma=6\pi/5$. }\label{fig2}
\end{figure}

 First, we consider  that the ideal conditions $\gamma=\pi$,
$t_a=t_b=t_1=t$  and $l_x=l_z=l$ are satisfied. In experiments,
these conditions can be achieved.  Taking the Fourier
transformation, $\hat a(\bk)=\sum_{(m(odd),n)} \hat
a_{m,n}\exp(-i\bk\cdot \br_{m,n})$ and $ \hat
b(\bk)=\sum_{(m(even),n)} \hat b_{m,n}\exp(-i\bk\cdot \br_{m,n})$,
we obtain the total Hamiltonian as
\begin{eqnarray}
\hat H&=&-2t\sum_k[\cos(k_zl)\hat b^\dag(\bk)\hat
b(\bk)-\cos(k_zl)\hat a^\dag(\bk)\hat a(\bk)\nonumber\\
&&+\cos(k_x l)\hat a^\dag(\bk)\hat b(\bk)+\cos(k_x l)\hat
b^\dag(\bk)\hat a(\bk)].\label{H3}
\end{eqnarray}
Diagonalizing the above Hamiltonian (\ref{H3}), we obtain the
quasiparticle energy spectrum    $ E(\bk)=
2st\sqrt{\cos^2(k_xl)+\cos^2(k_zl)} $ with $s=\pm 1$ being the band
index, which is similar to the spectrum of $\pi$ flux states in
quantum spin liquids \cite{Wen}. This energy spectrum has two energy
bands and
 contains four  zero-energy
Dirac points, where the conduction and
 valence bands intersect, in the first Brillouin zone at $ {\bf
K}_1=\left({\pi}/{2l},{\pi}/{2l}\right), {\bf
K}_2=\left(-{\pi}/{2l},{\pi}/{2l}\right), {\bf
K}_3=\left({-\pi}/{2l},-{\pi}{2l}\right), {\bf
K}_4=\left({\pi}/{2l},-{\pi}/{2l}\right)$.  Near the Dirac points,
the energy dispersion has standard cone-like shape as shown in
Fig.\ref{fig2} (a) and (b) and  the  spectrum is linear. The
low-energy state dynamics are described by
 linearizing their spectrum about the degeneracy points and are
 modeled by massless relativistic  fermions.

 For simplicity, we only consider the  part around the
Dirac points ${\bf K}_1$, and the physics around the other Dirac
points are similar. Setting $\bk={\bf K}_1+\bp$, we linearize the
Hamiltonian around the Dirac point ${\bf K}_1$ as, $ \hat H =\hbar
v_0\sum_p[p_z\hat b^\dag(\bp)\hat b(\bp)-p_z\hat a^\dag(\bp)\hat
a(\bp)+p_x\hat a^\dag(\bp)\hat b(\bp)
 +p_x\hat b^\dag(\bp)\hat a(\bp)]$ with $v_0=2tl/\hbar$,
which can be rewritten in coordinate space as, $ \hat H=\int d^2r
\hat\eta^\dag(\br)\hat{\cal H}\hat\eta(\br) $, where
$\hat\eta=(\hat\eta_{b}, \hat\eta_a)^T$ with $\hat\eta_b(\br)=\int
d^2pe^{-i\bp\cdot\br}\hat b(\bp)$ and $\hat\eta_a(\br)=\int
d^2pe^{-i\bp\cdot\br}\hat a(\bp)$. Here, $\hat{\cal H}$ is the
single-particle Hamiltonian as $\hat{\cal H}= \hbar
v_0(\hat{p}_x\sigma_x+\hat{p}_z\sigma_z)$, where $\sigma_x$ and
$\sigma_z$ are Pauli matrixes.  We obtain the eigenstates
\begin{eqnarray}
&&\phi_\bp^s=\frac{1}{\sqrt{2}}\left(\matrix{\cos\frac{\alpha}{2}+s\sin\frac{\alpha}{2}\cr
s\cos\frac{\alpha}{2}-\sin\frac{\alpha}{2}}\right)e^{{i}\bp\cdot\br},
\end{eqnarray}
where $s=\pm1$ and $\tan\alpha=p_z/p_x$. The corresponding
eigenenergies are $ E^s(\bp)=s \hbar v_0p $ with
$p=\sqrt{p_x^2+p_z^2}$. When the wave vector is
$\bp=p_x\hat{x}+p_z\hat{z}$, the corresponding group velocity and
pseudospin vector are ${\bf v}_g=sv_0(p_x\hat{x}+p_z\hat{z})/p$ and
${\bf c}=(p_x\hat{x}+p_z\hat{z})/p$, respectively. It is easy to
find that the three vectors ${\bf v}_g$, ${\bf c}$ and $\bp$ are
collinear, i.e., they are parallel to each other.  There is an
intimate relation between the pseudospin and motion of the
quasiparticle or quasihole: pseudospin can only be directed along
the propagation direction (say, for quasiparticles) or only opposite
to it (for quasiholes). As a result, quasiparticles or quasiholes
exhibit a linear dispersion relation $E=\hbar v_0 k $, as if they
were massless relativistic particles but the role of the speed of
light is played here by the Fermi velocity $v_0$.

In practice, the parameters may have fluctuations around the ideal
conditions considered above. Fortunately, even the parameters
deviate  from the ideal ones, the massless Dirac fermion spectrum
persists and remarkably exhibit anisotropic behaviors,  which are
just pursued in References \cite{Park1}  by adding external periodic
potentials on graphene. Here, we provide alternative methods to
exhibit anisotropic behaviors of massless  Dirac fermions in a
square optical lattice by setting the parameters deviated from the
ideal situation.

For simplicity, we only consider three cases with the existence of
parameter deviation from the ideal situation as follow: (i)
$t_a=t_b\neq t_1=t$,  $l_z=l_x=l$, $\gamma=\pi$; (ii)
$t_a=t_b=t_1=t$, $l_z\neq l_x=l$, $\gamma=\pi$; (iii)
$t_a=t_b=t_1=t$, $l_x=l_z=l$, $\gamma=\pi+\delta$ with $\delta\neq
0$. The corresponding dispersion relations are $E_{\rm
i}(\bk)=2st\sqrt{\cos^2(k_xl)+(t_a/t)^2\cos^2(k_zl)}$, $E_{\rm
ii}(\bk)= 2st\sqrt{\cos^2(k_xl_x)+\cos^2(k_zl_z)}$ and $E_{\rm
iii}(\bk)=t[\cos(k_zl+\delta)-\cos(k_zl)]+st\sqrt{4\cos^2(k_xl)+[\cos(k_zl+\delta)+\cos(k_zl)]^2}$
for cases (i), (ii) and (iii), respectively, which are shown in
FIG.\ref{fig2} (c)-(h). For case (ii), the four Dirac points are $
{\bf K}_1=\left({\pi}/{2l_x},{\pi}/{2l_z}\right), {\bf
K}_2=\left(-{\pi}/{2l_x},{\pi}/{2l_z}\right), {\bf
K}_3=\left({-\pi}/{2l_x},-{\pi}/{2l_z}\right), {\bf
K}_4=\left({\pi}/{2l_x},-{\pi}/{2l_z}\right)$, which are dependent
on the lattice spacing in the $x$ and $z$ direction, while the Dirac
points for cases (i)  are the same as those of the ideal case. For
case (iii), the four Dirac points are $ {\bf
K}_1=\left({(\pi-\delta)}/{2l},{\pi}/{2l}\right), {\bf
K}_2=\left({(-\pi-\delta)}/{2l},{\pi}/{2l}\right), {\bf
K}_3=\left({(-\pi-\delta)}/{2l},-{\pi}/{2l}\right), {\bf
K}_4=\left({(\pi-\delta)}/{2l},-{\pi}/{2l}\right)$. Around the Dirac
points, these spectra can be linearized as $E_{\rm i}^s(\bp)=s \hbar
v_0p_1$ with $p_1=\sqrt{p_x^2+f_1^2p_z^2}$, $E_{\rm ii}^s(\bp)=s
\hbar v_0 p_1$ with $p_2=\sqrt{p_x^2+f_2^2p_z^2} $ and $E_{\rm
iii}^s(\bp)=\pm 2t\sin(\delta/2)+s\hbar v_0p_3$ with
$p_3=\sqrt{p_x^2+f_3^2 p_z^2}$, where $f_1=t_a/t=t_b/t$, $f_2=l_z/l$
and $f_3=\cos(\delta/2)$. The corresponding single-particle
Hamiltonian can be written as $ \hat{\cal H}_{\rm i}= \hbar
v_0(\hat{p}_x\sigma_x+f_{1}\hat{p}_z\sigma_z) $, $ \hat{\cal H}_{\rm
ii}= \hbar v_0(\hat{p}_x\sigma_x+f_{2}\hat{p}_z\sigma_z) $ and
  $ \hat{\cal H}_{\rm iii}= \pm 2t\sin(\delta/2)+\hbar
v_0(\hat{p}_x\sigma_x+f_{3}\hat{p}_z\sigma_z) $ for cases
(i),(ii),(iii), respectively. In all   cases, the quasiparticles or
quasiholes are still massless Dirac fermions and  show chiral
behavior.  For the wave vector $\bp=p_x\hat{x}+p_z\hat{z}$, the
 group velocity are pseudospin vector are ${\bf
 v}_g=sv_t(p_x\hat{x}+f_j^2p_z\hat{z})/p_j$ and ${\bf
 c}=(p_x\hat{x}+f_jp_z\hat{z})/p_j$ for $j=1,2,3$. Here, the three vectors ${\bf v}_g, {\bf c}$ and $\bp$ are not
collinear and the dispersion relations near the Dirac points show
anisotropic behaviors.

\begin{figure}[ht]
\includegraphics[width=0.4\columnwidth]{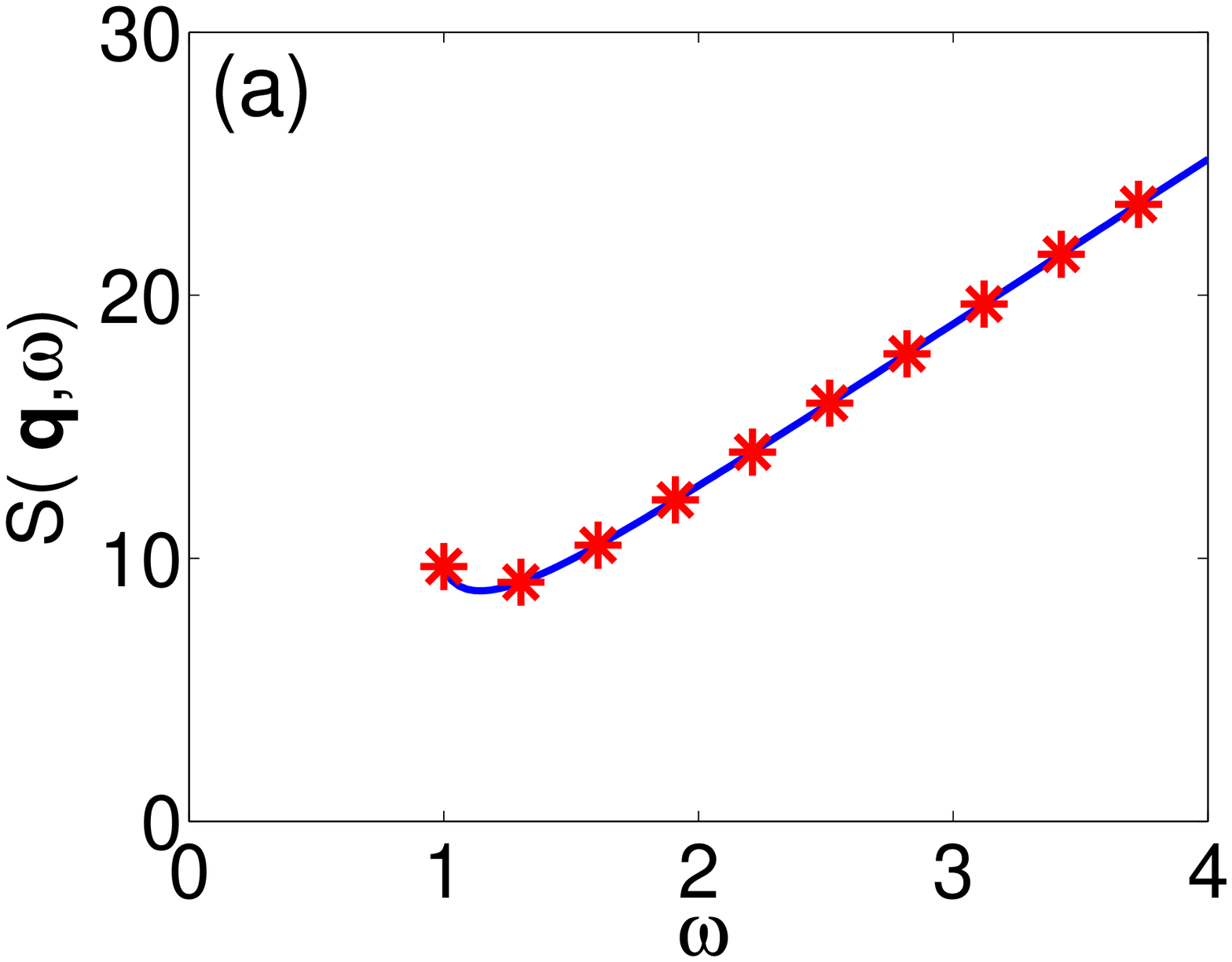}
\includegraphics[width=0.4\columnwidth]{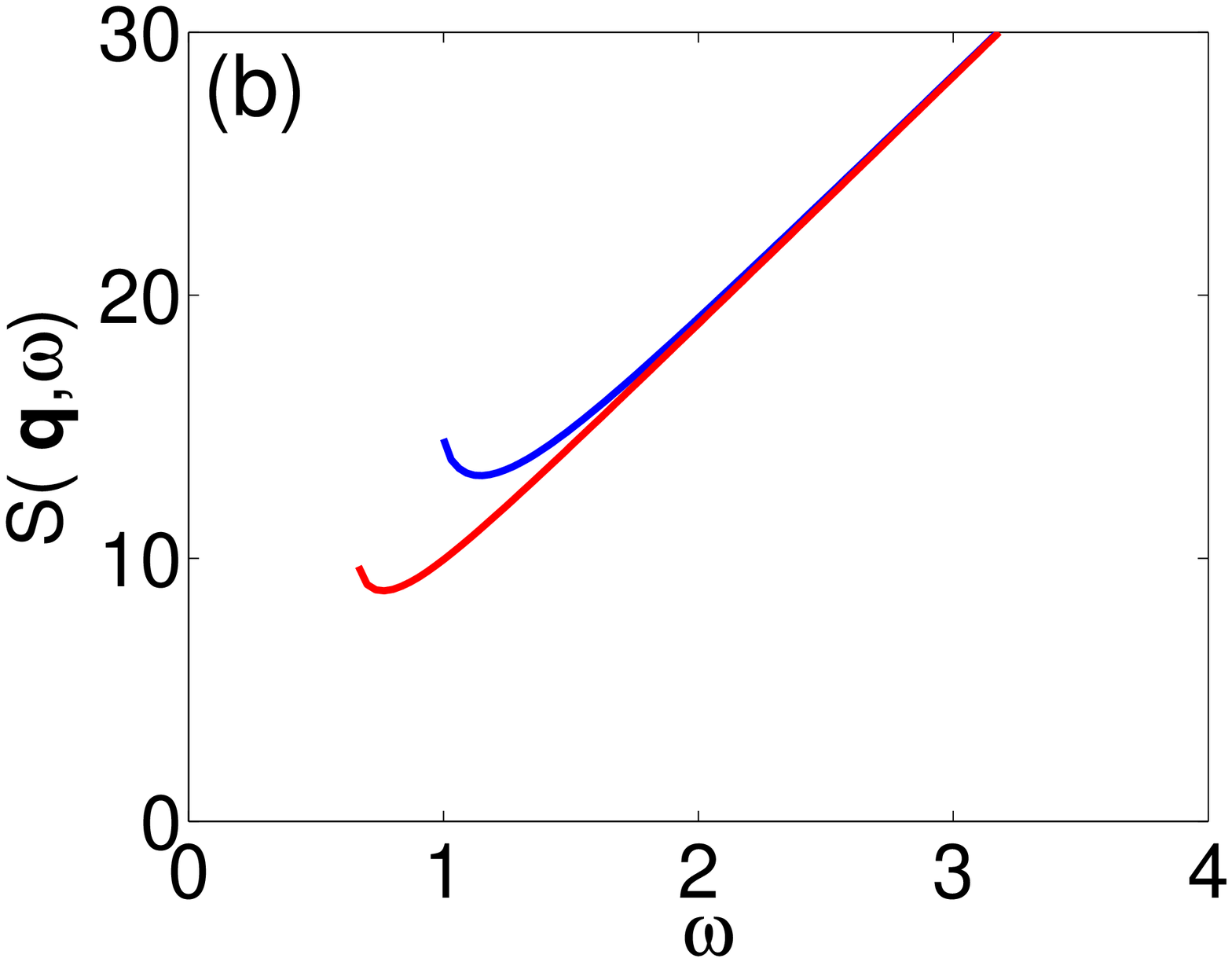}
\includegraphics[width=0.4\columnwidth]{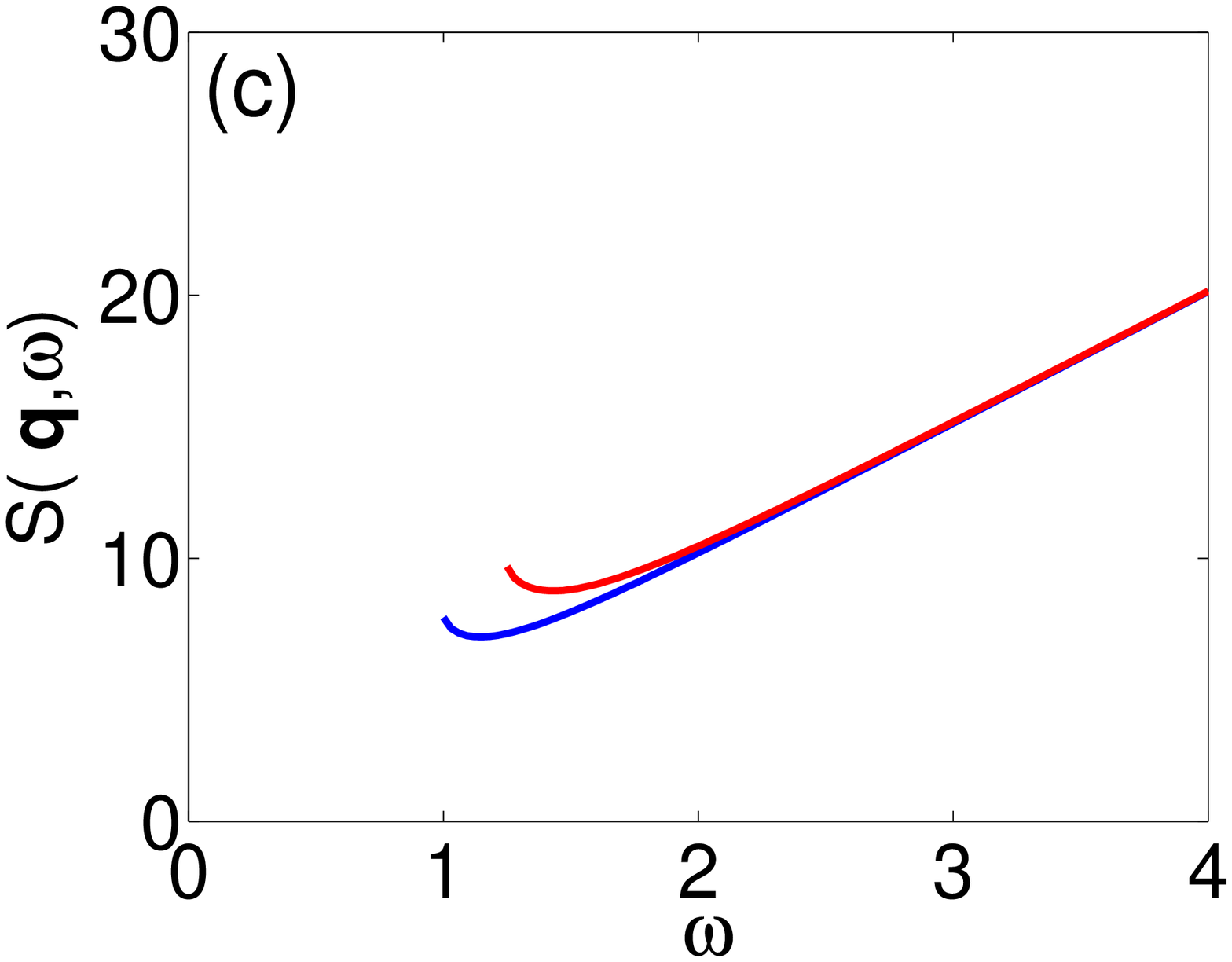}
\includegraphics[width=0.4\columnwidth]{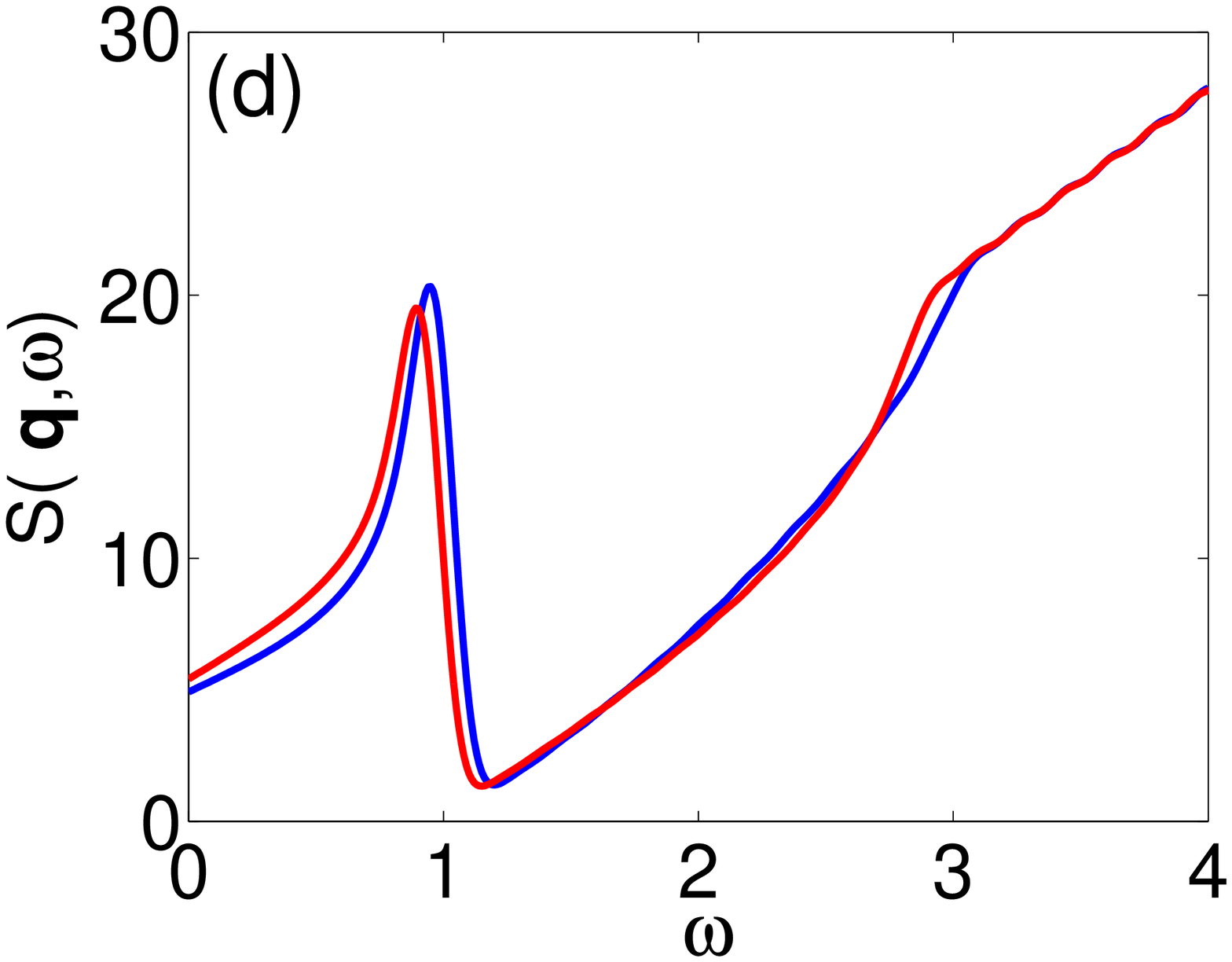}
\caption{The Bragg spectroscopies  for the ideal case,  (a)
$t_a=t_b=t_1=t$, $l_x=l_z=l$, $\gamma=\pi$, and the anisotropic
cases, (b) $t_a=t_b=2t_1/3=2t/3$,  $l_x=l_z=l$, $\gamma=\pi$, (c)
$t_a=t_b=t_1=t$, $l_z=5l_x/4=5l/4$, $\gamma=\pi$, (d)
$t_a=t_b=t_1=t$, $l_x=l_z=l$, $\gamma=6\pi/5$ and $q=\pi/10l$. Here,
we represent the Bragg spectroscopies with blue lines for the mentum
difference $\bq$  in the $x$ direction and with  red stars or red
lines for $\bq$ in the $z$ direction. The frequency difference
$\omega$ is expressed in units of $qv_0$ and the dynamic structure
factor $S(\bq,\omega)$ is expressed in units of $q/8\pi^2 n v_0$
with $n$ being the number density of atoms in the system.
}\label{Bragg}
\end{figure}

\section{Bragg spectroscopy}
Here, we propose  to  identify   massless Dirac fermionic
quasiparticles  with Bragg spectroscopy \cite{Stamper-Kurn}, which
is extensively used to probe  excitation spectra in condensed matter
physics. In Bragg scattering, the atomic gas is exposed to two laser
beams, with wavevectors $\bk_1$ and $\bk_2$ and a frequency
difference $\omega$. The light-atom interaction Hamiltonian for
Bragg scattering  can be written as, $ \hat
H_B=\sum_{\bp_1,\bp_2}\hbar\Omega_B e^{-i\bq\cdot
\br}|\phi_{\bp_2}^f\rangle\langle \phi_{\bp_1}^i|+{\rm H.c.} $
 with $\bq=\bp_2-\bp_1$, where the initial state $|\phi_{\bp}^i\rangle$ is a filled state under Fermi surface and
 the final state $|\phi_{\bp}^f\rangle$  is an empty state above Fermi surface. From the Fermi's golden rule, we obtain the
dynamic structure factor as follows,
\begin{eqnarray}
S(\bq,\omega)&=&\frac{1}{N\hbar^2\Omega_B^2}\sum_{\bp}|\langle
\phi_{\bp+\bq}^f|\hat H_B|\phi_{\bp}^i\rangle|^2\nonumber\\
&&\times\delta(\hbar\omega-E_{\bp+\bq}^f+E_{\bp}^i),
\end{eqnarray}
where $N$ is the total number of atoms in the system.

 Here, we consider  the case of half filling of cold fermions in the optical lattice, i.e. the
Fermi energy surface is at zero energy level, which is just at Dirac
points for the cases except anisotropic case (iii). The numerical
evaluation results of the dynamic structure factor is shown in FIG.
\ref{Bragg}. We note that there are lower cutoff frequencies
$\omega_r$ for the fixed momentum difference $q$ and $S(\bq,\omega)$
are approximately linear to the frequency difference $\omega$ for
large frequency difference $\omega$ for FIG.\ref{Bragg} (a),(b),
(c). However, for FIG.\ref{Bragg}(d), the cutoff disappears for the
Fermi surface is not at Dirac points for this case.  FIG.\ref{Bragg}
(a) show that the bragg spectroscopy curves for the momentum
difference $\bq$ in the $x$ and $z$ directions are identical in the
ideal case, which is just a consequence of the isotropy of the
energy spectrum. From  FIG.\ref{Bragg} (b), (c) and (d), we clearly
see  that the Bragg spectroscopies are different for $\bq$ in the
$x$ and $z$ directions in the anisotropic cases, which features the
anisotropic behaviors of those spectra.

\section{Conclusion}

In summary, we have proposed a novel scheme to realize massless
Dirac fermions in a 2D square optical lattice with assistance of
laser fields. For massless Dirac fermions, the gap is zero and the
linear dispersion law holds.  Our scheme is very robust against
perturbations. Even   the experimental situation deviates from the
ideal conditions, massless Dirac fermions persist and, furthermore,
exhibit novel features, i.e., anisotropic behaviors. Due to the
absence of hexagonal symmetry, our scheme suggests a new direction
to study Dirac fermions in the optical lattice.

\begin{acknowledgments}
This work was  supported  by the Teaching and Research Foundation
for the Outstanding Young Faculty of Southeast University. X. J. Liu
acknowledges support from US NSF Grant No. DMR-0547875  and  ONR
under Grant No. ONR-N000140610122.
\end{acknowledgments}

\end{document}